\documentclass[aps,prb,onecolumn,superscriptaddress,showpacs]{revtex4-2}
\usepackage{ulem}
\usepackage[utf8]{inputenc}
\usepackage{graphicx}
\usepackage{bm}
\usepackage{hyperref}
\usepackage{amsmath}
\usepackage{mathtools}
\usepackage[dvipsnames]{xcolor}

\usepackage{multirow}

\begin{document}
	
	\title{Impact of atomic defects in the electronic states of  FeSe$_{1-x}$S$_{x}$ superconducting crystals          }
	
	\author{Jazm\'{i}n Arag\'{o}n S\'{a}nchez}%
	\affiliation{Instituto de Nanociencia y Nanotecnología, CNEA and
		CONICET, Nodo Bariloche, Avenida Bustillo 9500, 8400 Bariloche, Argentina}
	\affiliation{Centro At\'{o}mico Bariloche and
		Instituto Balseiro, CNEA and Universidad Nacional de Cuyo, Avenida
		Bustillo 9500, 8400 Bariloche, Argentina}
	\affiliation{Leibniz Institute for Solid State and Materials Research, Helmholtzstra$\beta$e 20, 01069 Dresden, Germany}
	
	\author{Mar\'{i}a Lourdes Amig\'{o}}
	\affiliation{Instituto de Nanociencia y Nanotecnología, CNEA and
		CONICET, Nodo Bariloche, Avenida Bustillo 9500, 8400 Bariloche, Argentina}
	\affiliation{Centro At\'{o}mico Bariloche and
		Instituto Balseiro, CNEA and Universidad Nacional de Cuyo, Avenida
		Bustillo 9500, 8400 Bariloche, Argentina}
	
	\author{Cristian Horacio Belussi}
		\affiliation{Instituto de Nanociencia y Nanotecnología, CNEA and
		CONICET, Nodo Bariloche, Avenida Bustillo 9500, 8400 Bariloche, Argentina}
	\affiliation{Centro At\'{o}mico Bariloche and
		Instituto Balseiro, CNEA and Universidad Nacional de Cuyo, Avenida
		Bustillo 9500, 8400 Bariloche, Argentina}
	
	\author{Mar\'{i}a Victoria Ale Crivillero}
	\affiliation{Instituto de Nanociencia y Nanotecnología, CNEA and
		CONICET, Nodo Bariloche, Avenida Bustillo 9500, 8400 Bariloche, Argentina}
	\affiliation{Max-Planck-Institute for Chemical Physics of Solids, N\"{o}thnitzer Stra{\ss}e 40, 
		01187 Dresden, Germany}
	
	\author{Sergio Su\'{a}rez}
	\affiliation{Centro At\'{o}mico Bariloche and
		Instituto Balseiro, CNEA and Universidad Nacional de Cuyo, Avenida
		Bustillo 9500, 8400 Bariloche, Argentina}

	\author{Julio Guimpel}
	\affiliation{Instituto de Nanociencia y Nanotecnología, CNEA and
		CONICET, Nodo Bariloche, Avenida Bustillo 9500, 8400 Bariloche, Argentina}
	\affiliation{Centro At\'{o}mico Bariloche and
		Instituto Balseiro, CNEA and Universidad Nacional de Cuyo, Avenida
		Bustillo 9500, 8400 Bariloche, Argentina}
	
	\author{Gladys Nieva}
	\affiliation{Instituto de Nanociencia y Nanotecnología, CNEA and
		CONICET, Nodo Bariloche, Avenida Bustillo 9500, 8400 Bariloche, Argentina}
	\affiliation{Centro At\'{o}mico Bariloche and
		Instituto Balseiro, CNEA and Universidad Nacional de Cuyo, Avenida
		Bustillo 9500, 8400 Bariloche, Argentina}
	
	\author{Julio Esteban Gayone}
	\affiliation{Instituto de Nanociencia y Nanotecnología, CNEA and
		CONICET, Nodo Bariloche, Avenida Bustillo 9500, 8400 Bariloche, Argentina}
	\affiliation{Centro At\'{o}mico Bariloche and
		Instituto Balseiro, CNEA and Universidad Nacional de Cuyo, Avenida
		Bustillo 9500, 8400 Bariloche, Argentina}

	\author{Yanina Fasano}
	\email[]{Corresponding author: yanina.fasano@cab.cnea.gov.ar}
	\affiliation{Instituto de Nanociencia y Nanotecnología, CNEA and
	CONICET, Nodo Bariloche, Avenida Bustillo 9500, 8400 Bariloche, Argentina}
\affiliation{Centro At\'{o}mico Bariloche and
	Instituto Balseiro, CNEA and Universidad Nacional de Cuyo, Avenida
	Bustillo 9500, 8400 Bariloche, Argentina}
\affiliation{Leibniz Institute for Solid State and Materials Research, Helmholtzstra$\beta$e 20, 01069 Dresden, Germany}

	\vspace{10pt}
\date{\today}
	
	\begin{abstract}
		The electronic properties of Fe-based superconductors are drastically affected by
		deformations on their crystal structure introduced by doping and pressure. Here we study single crystals of  FeSe$_{1-x}$S$_{x}$ and reveal that local crystal deformations such as atomic-scale defects impact the spectral shape of the electronic core level states of the material.  By means of scanning tunnelling
		microscopy (STM) we image S-doping
		induced defects as well as diluted dumbbell defects associated with
		Fe vacancies. We have access to  the electronic structure of the samples by
		means of X-ray photoemission spectroscopy (XPS) and show that the
		spectral shape of the Se core levels can
		only be adequately described by considering a principal plus a
		minor component of the electronic states. We find this
		result for both pure and S-doped samples, irrespective  that in the latter case the material presents extra crystal defects associated to doping with S atoms. We argue that the second
		component in our XPS spectra is associated with the ubiquitous
		dumbbell defects in FeSe that are known to entail a significant modification of the electronic clouds of surrounding atoms.

	\end{abstract}
	
\maketitle	
	
	\section*{Introduction}
	
	The so called iron age of high-temperature superconductivity renewed the hope on discovering 
	superconducting compounds with technically relevant critical temperatures. This age
	was inaugurated by the discovery of superconductivity with a critical temperature $T_{\rm c}=26$\,K in the 
	superconductor LaO$_{1-x}$F$_{x}$FeAs~\cite{Kamihara2008} and was followed
	by numerous reports of superconductivity in  the
	Fe-based superconductors family.~\cite{Hsu2008,Rotter2008,Tapp2008,Pitcher2008,Wang2008} Among all the members of this
	family, the  FeSe compound~\cite{Hsu2008} has attracted much attention since
	it has a simple crystal structure of stacks of superconducting layers and its $T_{\rm c}$ 
	can be enhanced up to 37\,K by applying hydrostatic pressure,~\cite{Margadonna2009,Medvedev2009} the largest value for a binary compound.

	In addition, FeSe is special since puzzles the understanding of
	the nature of high-temperature superconductivity in Fe-based superconductors: While  most of 
	them presents a magnetic order that seems to be in close
	relationship with superconductivity,~\cite{Norman2008,Buchner2009} in FeSe no static magnetic order is observed at ambient pressure.~\cite{Buchner2009,Bohmer2018,Chen2020} 
	This compound also presents a tetragonal to orthorhombic transition on cooling around 90\,K~\cite{Margadonna2008} without undergoing any magnetic transition.~\cite{McQueen2009,Bendele2010}   Another
	relevant part of the puzzle is that doping FeSe with  chalcogen atoms  or applying pressure
	produces significant changes in the phase diagram of the compounds~\cite{Watson2015} and enhances magnetic instability.~\cite{Imai2009}  Indeed, spectroscopic data obtained by means of scanning tunneling microscopy (STM) give support to the idea that spin fluctuations have a relevant role in the superconducting pairing mechanism of Fe-based superconductors.~\cite{Fasano2010,Hoffman2011,Chi2012} 
	All these results indicate that in FeSe there is an intricate interplay between the crystal structure
	and electronic properties such as superconductivity and magnetism.

	For instance, introducing chemical pressure in the material by
	isovalently substituting Se with another chalcogen element is a
	suitable control parameter to tune crystal structure
	deformations~\cite{Hanaguri2018} as well as the critical
	temperature.~\cite{Watson2015} Other examples of the fine interplay
	between crystal structure and electronic properties in FeSe is the
	radical enhancement of $T_{\rm c}$   when growing strained monolayer
	films on top of SrTiO$_{3}$ substrates~\cite{Ge2014}, or when
	coating the samples with K adatoms.~\cite{Miyata2015} Atomic-scale
	local crystal structure modifications such as defects do also play a
	role in the electronic and superconducting properties of the
	material.~\cite{Kamihara2008,Yeh2008} A  prominent example of this
	is reported in molecular beam epitaxy (MBE)-grown FeSe films
	presenting dumbbell-type atomic defects: Superconductivity is
	suppressed on tuning the defect concentration above a relatively low
	critical value.~\cite{Song2011} Thus, understanding the impact of
	atomic-scale defects in the electronic structure of Fe-based
	superconductors is of key importance for assessing how critical is
	the occurrence  of these features for the establishment of
	superconductivity in these compounds.

	With the aim of studying this impact, here we study single crystals
	of FeSe$_{1-x}$S$_{x}$, the Fe-based compound with the simplest
	crystal structure, consisting of a stack of superconducting layers.
	We apply scanning tunnelling microscopy (STM) to reveal that this
	compound presents S-doping induced defects as well as diluted
	dumbbell defects associated with an Fe vacancy. We measure the
	electronic structure of the samples by means of X-ray photoemission
	spectroscopy (XPS) and reveal that the spectral shape of the peaks
	of some of the Se and Fe core levels can only be adequately
	described by considering a dominant plus a smaller second component
	of the electronic states. We find this result for both, pure as well
	as S-doped samples, irrespective that they present extra crystal
	defects associated with the substitution of Se by S atoms.
	Structural and resistivity characterization, as well as the spectral
	shape and energy location of the peaks in XPS spectra, indicate our
	samples do not present inter-growths of the hexagonal phase. Even
	though in our STM topographies only $\sim 4$\,\% of the imaged Se
	atoms are involved in dumbbell defects, according to DFT
	calculations these defects entail a significant modification of the
	electronic clouds of the eight Se and Fe atoms surrounding the Fe
	vacancy. Thus, we suggest the second component in our XPS spectra is associated
	with the ubiquitous dumbbell defects in FeSe. This impact of the atomic
	defects in the binding energy and the spectral shape of the core levels
	in FeSe$_{1-x}$S$_{x}$ highlights the subtle interplay between the
	crystal structure and the bulk electronic states in Fe-based
	superconductors.

	\section*{Experimental details and characterization of samples}
	
	In this work we study platelet-like FeSe$_{1-x}$S$_{x}$ single
	crystals obtained by means of the vapor transport method.  Crystals
	were grown during 45 days using a KCl: 2AlCl$_{3}$ flux and a
	temperature-gradient with a hot-point of 395\,$^{\circ}$C and a cold-point of
	385\,$^{\circ}$C. Further details on the crystal growing method  can be
	obtained in Ref.\,\cite{Amigo2014}. X-ray diffraction experiments at
	room temperature indicate the single crystallinity and the
	tetragonal structure of the samples and do not  show any detectable
	trace of spurious phases.~\cite{Amigo2014}  We
	have not detected a significant concentration of Al nor K neither
	with energy-dispersive X-ray spectroscopy nor with time-of-flight
	secondary ion mass spectrometry techniques. The XPS spectra measured
	in our samples show no peaks that could be associated with the K 2p
	doublet nor Al 2s and Al 2s levels. Thus, no significant
	concentration of Al and K from the growing flux is detected in our
	samples.

\begin{figure}[ttt]
		\centering
		\includegraphics[width=\linewidth]{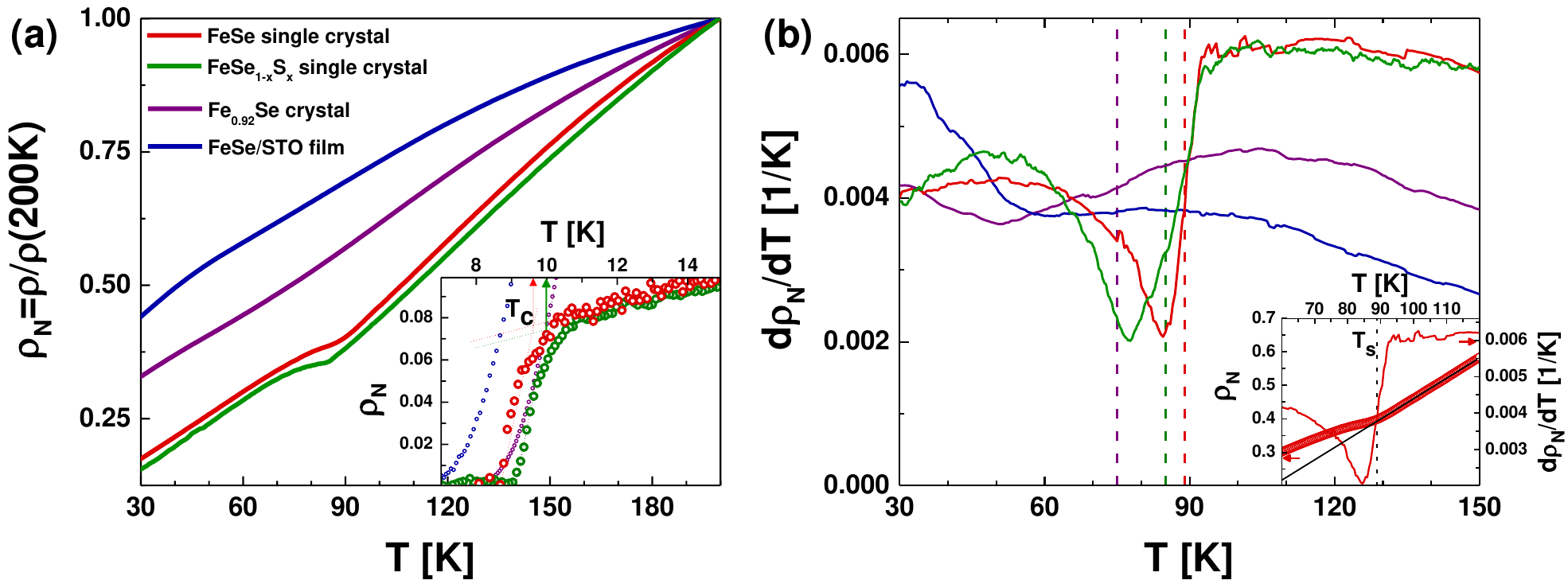}
		\caption{Transport properties of FeSe$_{1-x}$S$_{x}$ single crystals. (a) Normalized resistivity $\rho_{\rm N}=\rho(T)/\rho(200\,K)$ in the intermediate temperature range for the pure and S-doped single crystals studied in this work (red and green points). We also compare data in single crystals with data in a FeSe crystal with inter-growths of spurious phases (violet points) and with data in a FeSe/STO film with important mosaicity (blue points). Insert: Detail of the superconducting transition of the single crystals at low temperatures with the superconducting critical temperature $T_{\rm c}$ indicated with arrows. (b) Derivative of  the normalized resistivity with respect to temperature for the curves in panel (a). The tetragonal to orthorhombic transition temperature $T_{\rm s}$ is indicated with dashed vertical lines. Insert: Criteria used to estimate $T_{\rm s}$ from the resistivity  and its derivative. }
		\label{fig:figure1}
	\end{figure}

	Figure\,\ref{fig:figure1} (a) presents the $ab$-plane normalized
	resistivity, $\rho_{\rm N}=\rho(T)/\rho(200\,\rm K)$, of the studied
	single crystals. As shown in the insert, the single crystals present
	superconducting transitions with $T_{\rm c}=9.6(0.2)$\,K for $x=0$,
	and 10(0.2)\,K for the S-doped sample, both with transition width of 2\,K. An increase of the
	superconducting critical temperature is expected in the latter due to the positive
	chemical pressure introduced by the S-dopant
	atoms.~\cite{Watson2015}
	
	Data in Fig.\,\ref{fig:figure1} (a) show that the resistivity
	$\rho_{\rm N}$ of the crystals present a kink at an intermediate
	temperature. This feature is associated with the structural transition
	from tetragonal to orthorhombic symmetry on cooling, occurring at
	$T_{\rm s}\sim 90$\,K for pure FeSe according to elastic neutron
	scattering measurements.~\cite{Wang2016} This transition temperature
	$T_{\rm s}$ is manifested as a departure of the resistivity from its
	high-temperature linear behavior.~\cite{Wang2016} This temperature coincides
	with the midpoint of the jump in $d\rho_{\rm N}(T)/dT$ as indicated
	in the insert to Fig.\,\ref{fig:figure1} (b). Following this
	criteria we found $T_{\rm s} = 89 (1)$\,K for our $x=0$ and
	85(1)\,K for our S-doped single crystals, see dashed lines. This last value is in
	agreement with a doping range  $x= 0.02-0.03$ according to the
	literature.~\cite{Watson2015}

	We would like to point out that the temperature location and shape of the kink in the derivative of $d\rho_{\rm N}(T)/dT$ associated with the tetragonal to orthorhombic transition seems to be a
	good criteria to ascertain the purity and crystallographic
	quality of FeSe$_{1-x}$S$_{x}$ samples.~\cite{Wang2016,Bohmer2018}
	For comparison, in
	Fig.\,\ref{fig:figure1} we show data for a crystal from another
	batch  with formula Fe$_{0.92}$Se~\cite{Amigo2014} and for a
	sputtered FeSe/SrTiO$_{3}$ film,~\cite{AleCrivillero2019} both presenting inter-growths of the
	magnetic hexagonal phase. This film has a tweed grain
	pattern with a typical length scale $\sim 1$\,$\mu$m. In the case of
	the Fe-deficient crystal the kink in resistivity and jump in its
	derivative is fainted and a $T_{\rm s} \sim 75$\,K can be estimated.
	This decrease in the structural transition temperature is also found
	in Se-deficient crystals.~\cite{Margadonna2008} In the case of the
	FeSe film with strong mosaicity the minimum in $d\rho_{\rm N}(T)/dT$
	is not even evident.  Thus, we
	argue that the single crystalline samples we study in this work
	have a good crystal quality as ascertained by transport
	measurements.

\begin{figure}[ttt]
		\centering
		\includegraphics[width=0.6\linewidth]{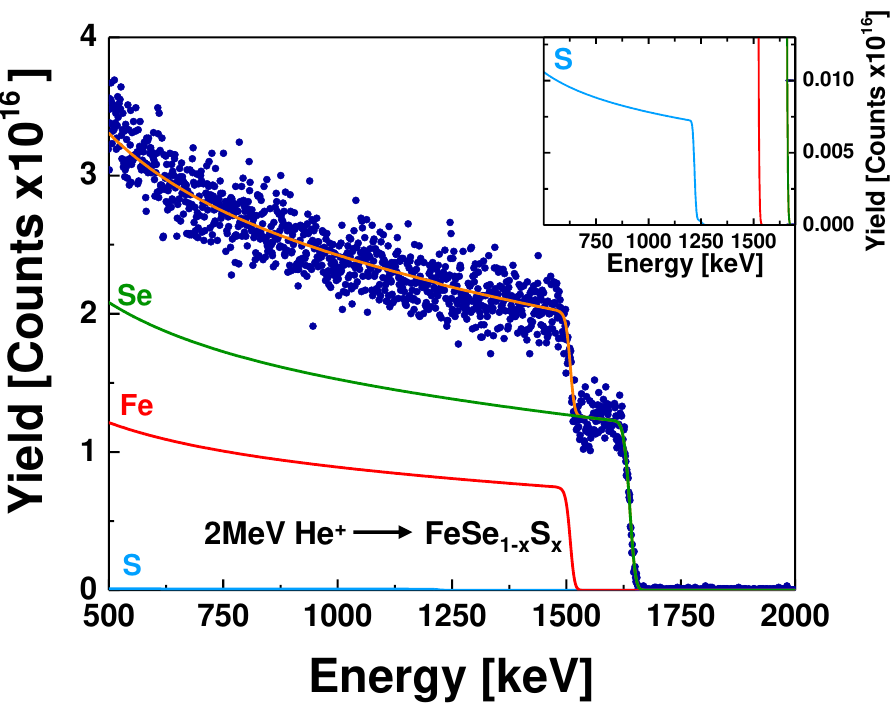}
		\caption{Rutherford backscattering spectrometry (RBS) data obtained at room temperature in a FeSe$_{1-x}$S$_{x}$ single crystal (blue points). The fit of the data with the SIMNRA code (orange line) and the individual Fe, Se and S contributions are indicated.
			Insert: Zoom in to the S contribution to the total fit.}
		\label{fig:figure2}
	\end{figure}

	We measured the level of S-doping in our samples by means of bulk
	and surface sensitive techniques. In the first case, X-ray
	energy-dispersive spectroscopy yields a bulk composition of $x=0.03
	(0.01)$. The level of S was also estimated by means of Rutherford
	backscattering spectrometry (RBS) measurements 
	performed using a 1.7 MV Tandem particle accelerator and irradiating the
	sample with 2\,MeV He ions. This technique is sensitive to the S
	composition in a surface layer of roughly 1\,$\mu$m.  The RBS data
	presented in Fig.\,\ref{fig:figure2} are  fitted with the SIMNRA code
	developed for the simulation of RBS and other spectra.~\cite{SIMNRA} The total fit of the
	spectra is shown with an orange line and the contributions of the
	different elements are indicated with colour lines. This fitting
	yields a S doping level  of $x= 0.04 (0.02)$. We also
	estimated the local concentration of S at the top layer of in-situ
	cleaved samples by means of STM. We
	found in the exposed surfaces an average value of $x=0.027 (0.005)$.

	In order to study the electronic states in our FeSe$_{1-x}$S$_{x}$
	single crystals, we applied STM and XPS techniques. STM data were obtained with a variable-temperature scanning tunnelling  microscope from
	Omicron Nanotechnology allowing cooling the sample down to $\sim
	30$\,K by means of a cold-finger connected to a continuous He-flux
	refrigeration system. Topographies were obtained in constant-current
	mode with tunnelling regulation conditions in the ranges of
	0.2-0.8\,V and 0.7-1.3\,nA. Typically, we observed images with
	atomic resolution when scanning at a speed between 30 and 40\,nm/s.
	XPS measurements were performed in a system for surface analysis
	from SPECS. The main chamber, with a base pressure in the low
	$10^{-10}$\,Torr range, is equipped with a high-resolution energy
	analyzer and a  monochromatic $Al_{K_{\alpha}}$ X-ray  source
	(1486.4 eV). The spot size of the XPS system is of few millimeters
	and the samples have 1\,mm$^2$ of area at maximum. Thus we are
	inevitably collecting signal coming from the sample holder and
	epoxy used to glue the sample.
	
	\section*{Results}

	\begin{figure}[ttt]
		\centering
		\includegraphics[width=\linewidth]{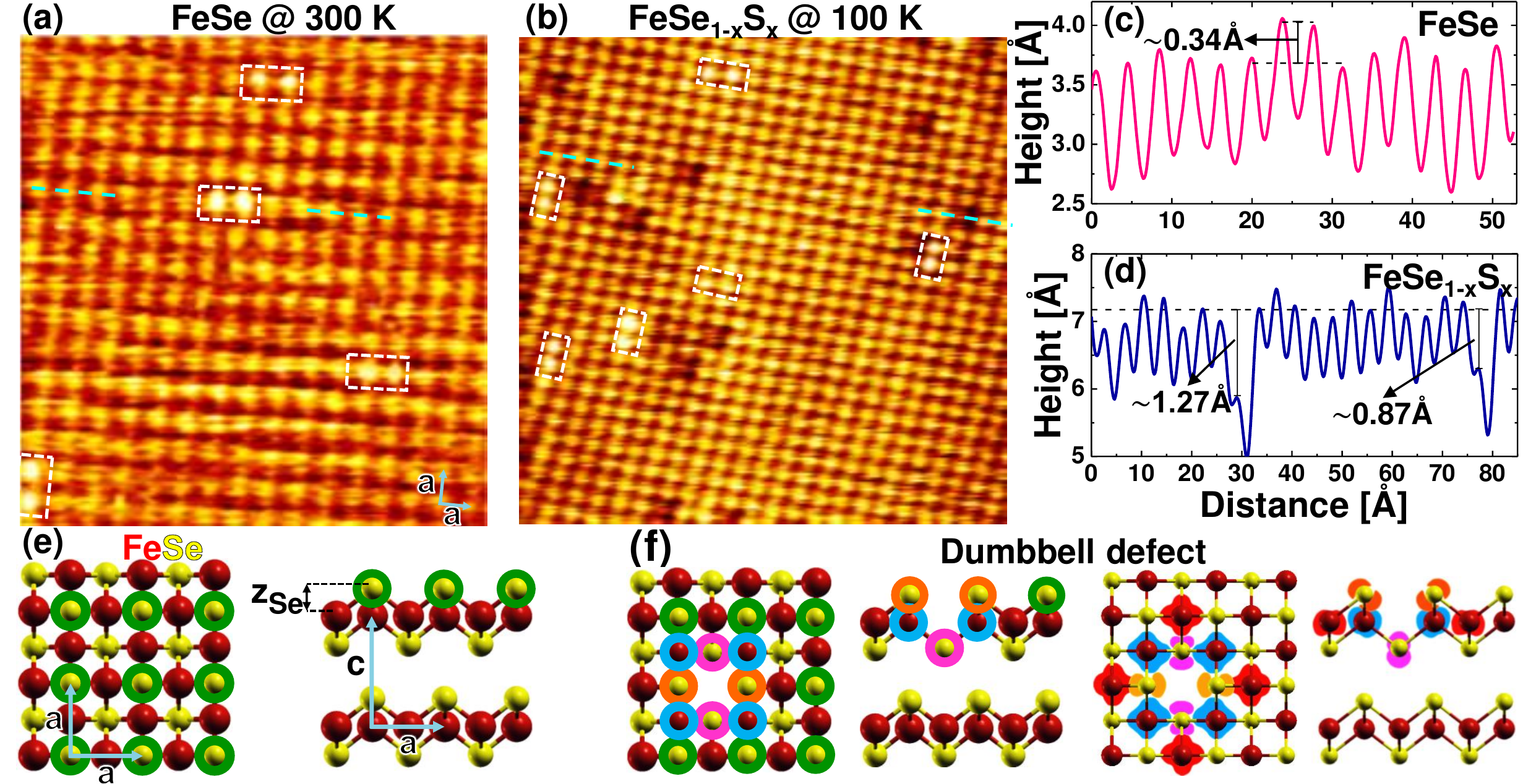}
		\caption{STM topography images and crystal structure of FeSe$_{1-x}$S$_{x}$
			in the high-temperature tetragonal phase. (a) $6\,\times\,6$\,nm$^2$ topography of an in-situ cleaved single crystal with $x=0$. The exposed Se atoms are observed as bright spots. Dumbbell defects (see white dashed frames) are observed as two bright Se atoms oriented along the $\mathbf{a}$-axis directions. Image acquired in constant-current mode at 0.7\,nA and 0.7\,V. (b) $10\,\times\,10$\,nm$^2$ topography of an in-situ cleaved crystal with $x=0.027 (0.005)$ at the surface.  Local depletions of the sample height (darker chalcogen atoms) are presumably generated by smaller S dopant atoms. Dumbbell defects are also indicated with dashed-white frames. Image acquired in constant-current mode at 1.3\,nA and 0.2\,V.  For both samples the measured lattice spacing is $a=3.8 (0.1)$\,\AA. (c) Height-profile in FeSe along the trace indicated (partially) with a turquoise dashed line in panel (a). A dumbbell defect is observed as two local maxima of apparent height due to the protrusion of the Se electronic cloud entailed in this defect.~\cite{Huang2016} (d) Height-profile in the S-doped crystal along the trace indicated with the turquoise line in panel (b). Local minima in the surface height are detected in the darker height features.  (e) Schematic representation of the FeSe crystal structure in the tetragonal phase. The expected $a=3.77$\,\AA\ and $c=5.52$\,\AA\ unit cell vectors are indicated. When cleaving the samples to perform STM measurements, the Se atoms located above the top Fe plane are imaged (highlighted in green). (f) Schematic of a dumbbell defect associated with an Fe vacancy: Atomic positions (left) and charge density isosurfaces (right) of the atoms entailed in the defect. The atomic clouds of the neighbouring Se and Fe atoms are schematically reproduced from the DFT calculations of Ref.~\cite{Huang2016}.}
		\label{fig:figure3}
	\end{figure}

	Figures \ref{fig:figure3} (a) and (b) show examples of STM
	current-constant topography images of the UHV in-situ cleaved
	FeSe$_{1-x}$S$_{x}$ crystals studied in the tetragonal phase (normal
	state). The intensity of every pixel of this image entails local
	information on the electronic properties of the material since
	corresponds to an integration of the local electron density of states of the
	sample up to the regulation voltage of the junction. The images present
	bright spots arranged in a two-dimensional square lattice. The
	crystal structure of the FeSe system is composed of a stack of
	Fe planes with Se atoms located a distance $z_{\rm Se} =
	1.47$\,\AA~\cite{McQueen2009} above and below the plane, see
	schematic of Fig.\,\ref{fig:figure3}. The easy cleaving plane of the
	samples is located between two adjacent Se planes. Thus, when
	cleaving the sample, the Se atoms located above the top Fe plane are
	exposed, see atoms highlighted in green in the schematic crystal
	structure of Fig.\,\ref{fig:figure3}\,(e). Therefore the bright
	spots in the topographic image correspond to Se atoms spaced a
	distance  $a$  for the tetragonal phase, see arrows in panel (a).
	Both in the case of pure or S-doped single crystals this separation
	is in average $3.8 (0.1)$\,\AA, in agreement with the value reported
	in the literature for the lattice spacing
	$a=3.77$\,\AA.~\cite{Margadonna2008}

	The topographic images for both types of samples show a ubiquitous
	feature: Pairs of brighter Se atoms aligned in the $\mathbf{a}$
	directions indicated with dashed-white frames in  Figs.\,\ref{fig:figure3} (a)
	and (b). These brighter atoms are observed as higher atomic peaks in
	traces of profile height as the one shown in Fig.\,\ref{fig:figure3}
	(c). Statistics in all our topographic images indicate the height
	difference in brighter atomic peaks with respect to the surrounding
	atoms is of $0.4 (0.1)$\,\AA\,in pure FeSe and $0.4 (0.3)$\,\AA\,in
	S-doped samples. The distribution of these features is quite
	diluted, representing $4.2 (0.6)$ and $3.8 (0.6)$\,\% of the STM-imaged Se atoms in the pure and S-doped samples respectively. Statistics were performed considering 1500 atoms in the case of pure FeSe and 5000 atoms in the case of S-doped samples. These features were also
	reported in STM topographic studies of in-situ MBE-grown FeSe
	films~\cite{Huang2016} and are also observed in studies of FeSe
	crystals by other authors.~\cite{Putilov2019} According to a previous report in the
	literature, in-situ grown FeSe films are no longer superconducting
	if a large density of Se atoms imaged at the sample surface are
	entailed in these features.~\cite{Huang2016}

	\begin{figure*}[ttt]
		\centering
		\includegraphics[width=\linewidth]{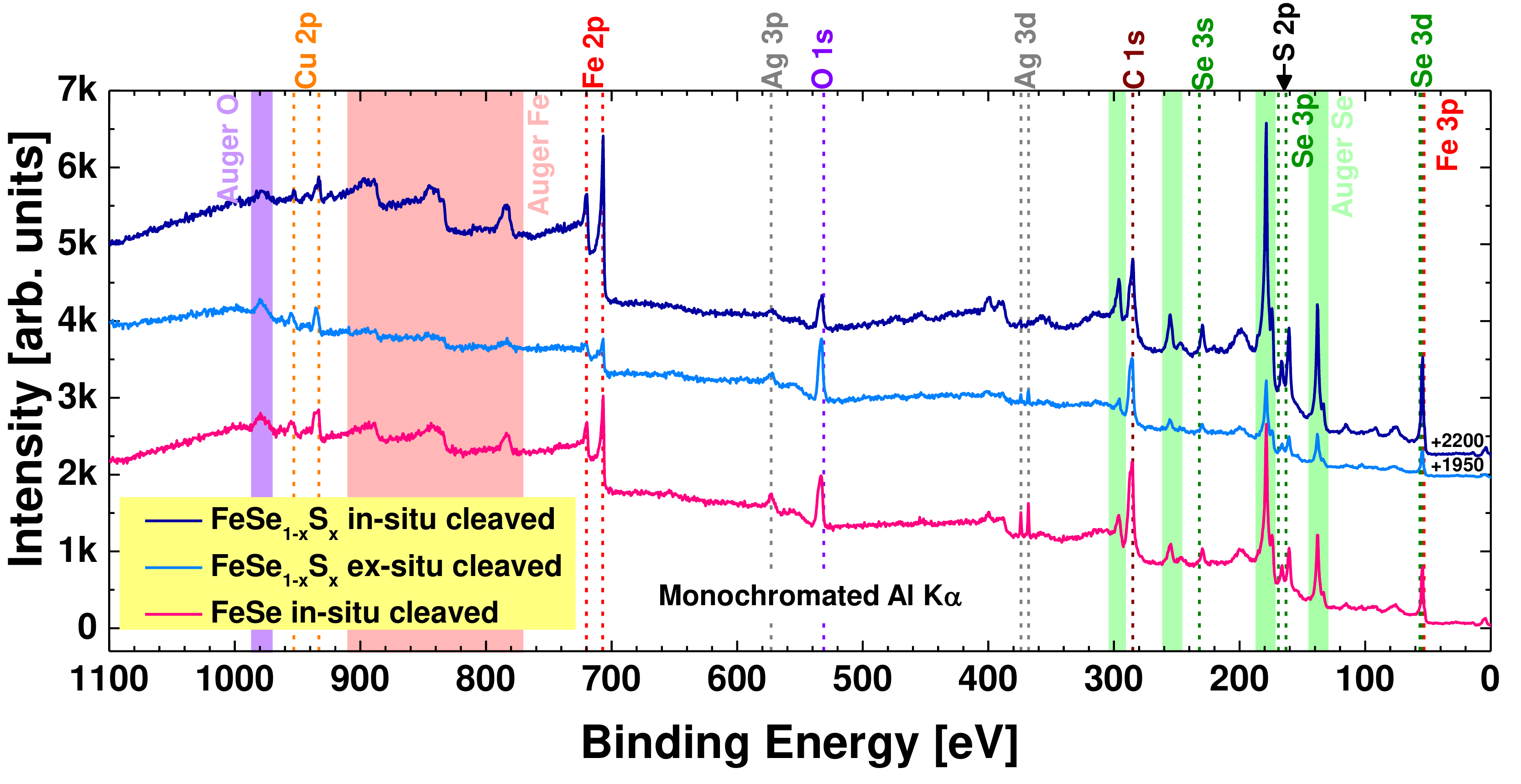}
		\caption{XPS spectra in an extended binding energy range for
			FeSe$_{1-x}$S$_{x}$ (blue in-situ and turquoise ex-situ cleaved) and
			FeSe (pink in-situ cleaved) samples. The measurements were performed
			using 1486.6\,eV  X-rays from the  K-$\alpha$ line of Al. Some spectra are vertically shifted for clarity (see labels in the right). The tabulated energy-location of the Fe 2p and Fe 3p  as well
			as the Se 3s, Se 3p and Se 3d core levels in pure elements are indicated with dashed
			lines; the  S 2p level is indicated with an arrow  (see labels at the top). The spurious peaks arising
			from the contribution
			of the sample holder (Cu 2p), conductive epoxy used to glue
			the samples (Ag 3p and Ag 3d) and
			contamination of the surfaces (C 1s and O 1s) are indicated with
			dashed lines. The energy regions where Auger structures for Fe, Se and O are expected are highlighted in red, green and violet, respectively. No peaks that could be associated with a significant concentration of Al and K from the growing flux are detected in these XPS spectra.}
		\label{fig:figure4}
	\end{figure*}

	By means of density functional theory simulations, the authors of Ref.\,\cite{Huang2016} showed that this brighter atoms are dumbbell defects associated with an Fe vacancy in the site in between the two brighter Se atoms. The schematic of Fig.\,\ref{fig:figure3} (f) shows that this defect generates a modification of the electronic cloud of the 2 Se atoms located above the plane and imaged as brighter by STM (orange), the 2 Se atoms located below (magenta), and the 4 first-neighbor Fe atoms (turquoise). The schematic of the charge density isosurfaces in the plane, shown in the right column of the figure, highlights the asymmetry in the shape of the electronic cloud of these 8 atoms with respect to the symmetric ones expected for atoms located further away from the defect (4 Fe atoms with red clouds). Furthermore, the same calculations show that the missing Fe atom induces that the orbitals on the 2 Se atoms located above the plane (orange ones) protrude further out than in the case of the other Se atoms (green). Thus, since a topographic image is proportional to the integral of the local density of states up to the regulation voltage, this protrusion of the electronic cloud results in an apparent larger height (brighter spots) of the Se atoms entailed in the dumbbell defect. Even though the density of dumbbell defects imaged by STM is small, these are defects imaged only at the surface, and such defects can certainly occur in every FeSe plane of the crystal. Thus, the concomitant modification of the electronic cloud of the 8 atoms directly involved in dumbbell defects might have a small though noticeable impact in the bulk electronic properties of the samples.

	In the case of the S-doped samples, another prominent feature is also
	imaged in several locations: Local  depletions of the
	sample height associated to darker chalcogen atoms. Height profiles along these features indicate that in
	the darker areas there is a height depletion of roughly 25\,\% with respect
	to neighbor Se atoms,  a magnitude of $\sim 20$\,\% the $c$-axis
	unit cell of the material, see for example Fig.\,\ref{fig:figure3}
	(d). These features are not detected in measurements in pure FeSe
	crystals performed by us  nor in works of other
	authors.~\cite{Hanaguri2018,Huang2016,Putilov2019} This type of
	defects were also observed in STM topographies of S-doped crystals
	with an occurrence that grows with the S
	concentration.~\cite{Hanaguri2018} Given that S has a smaller atomic
	radius than Se, it can be assumed that S atoms are imaged by STM as
	these darker features entailing a local depletion of height. Indeed,
	modulations of the brightness of topographic images are also
	observed in the FeTe$_{0.55}$Se$_{0.45}$ compound. In this work the
	darker spots are associated with the smaller atoms and the brighter
	ones with the larger atoms.~\cite{He2011} In view of this
	experimental evidence, our topographical data are consistent with
	the presence of S dopant-atoms in the darker chalcogen atom
	locations. Our STM results indicate that crystal disorder is more
	important in S-doped samples than in pure ones.

	We pursue our study of the electronic properties of
	FeSe$_{1-x}$S$_{x}$ by means of XPS measurements that provide
	information on the energy spectrum of the core levels of the
	different  elements composing the material. XPS is a surface
	sensitive technique, but in contrast to STM, the electronic
	information comes not only from the last atomic layer but also from the
	few atomic layers spanning a depth of $\sim 1$\,nm. Figure\,\ref{fig:figure4}
	shows the survey spectra of FeSe in-situ cleaved and
	FeSe$_{1-x}$S$_{x}$ ex and in-situ cleaved crystals. The ex-situ
	cleaved sample was cleaved in air right before putting it inside the
	preparation chamber. The spectra were obtained with an energy
	spacing of 0.5\,eV. The collected spectra in the FeSe$_{1-x}$S$_{x}$
	samples were vertically shifted for clarity. In the case of pure
	FeSe, our data are obtained in an in-situ cleaved surface of a high purity single crystal and cover a wider energy range with better
	signal to noise ratio and/or better energy resolution than
	previously reported XPS
	spectra.~\cite{Feng2004,Wu2007,Yoshida2009,Chen2010,Tsukada2011,Qi2011,Telesca2012,Chaluvadi2021}
	In all  spectra of Fig.\,\ref{fig:figure4}, Fe and Se contributions
	are  detected at energies close to the ones  corresponding to the
	tabulated peaks for the different core levels for the pure elements
	(indicated with dotted lines).~\cite{Moulder1992} Nevertheless, the
	S peaks are not clearly observed since they are superimposed to the Se
	3p peaks (see discussion below). Spurious Cu, Ag, O and C peaks are
	also detected due to the fact that the samples are smaller than the
	analyzed area, and we are detecting signals from the Cu sample
	holder and the conductive epoxy used to glue the sample.
	Table\,\ref{tab:comparisson} shows a comparison of the energy location of the
	Fe and Se core levels tabulated for pure elements, measured in our
	in-situ cleaved FeSe$_{1-x}$S$_{x}$ single crystals and other data available
	in the literature for pure FeSe crystals and films.
	
	\begin{table}[ttt]
		\centering
		\begin{tabular}{ |c||c|c|c|c|c|c|}
			\hline
			\multirow{3}{*}{Species} & \multicolumn{6}{|c|}{Binding Energy [eV]} \\
			\cline{2-7}
			& Pure & FeSe  & FeSe$_{0.97}$S$_{0.03}$ & FeSe  & FeSe/MBE  & FeSe/PLD \\
			& elements & this work  &  this work &  crystal \cite{Yamasaki2010} &  film \cite{Chen2010} &  film \cite{Chaluvadi2021} \\
			\hline
			Fe2p$_{3/2}$ & 707 & 706.5 & 706.5 & 706.9 &706.8 & 707.2  \\
			\hline
			Fe2p$_{1/2}$ & 720.1 & 719.8 & 719.8 & 719.6 & 720.1 & 720.7   \\
			\hline
			Fe3p$_{3/2}$ & 53 & 52.6 & 52.61 &  & 52.4 & 54.9   \\
			\hline
			Fe3p$_{1/2}$ & 53.67 & 53.25 & 53.23 &  &  & 56.6  \\
			\hline
			Se3p$_{3/2}$ & 163 & 160.9 & 160.95 &  & &  \\
			\hline
			Se3p$_{1/2}$ & 169 & 166.6 & 166.7 &  & &   \\
			\hline
			Se3d$_{5/2}$ & 55.6 & 54.15 & 54.15 & 54 & 54 & 54.48   \\
			\hline
			Se3d$_{3/2}$ & 56.46 & 55.05 & 55.05 & 54.9 & 55 & 55.65   \\
			\hline
		\end{tabular}
		\caption{Binding energies for the Fe and Se core levels. The first column has values tabulated for pure elements~\cite{Moulder1992} but in the case of Fe3p$_{1/2}$ where the value was calculated considering the binding energy of Fe3p$_{3/2}$ and the spin-orbit coupling. The second and third columns indicate the energy location of the peaks measured in our
			high purity in-situ cleaved FeSe$_{1-x}$S$_{x}$ crystals. The uncertainty in the data is of 0.05\,eV. We also include our
			estimation (from data in the figures of the corresponding papers) of
			the energy location of the peaks for the FeSe single crystals of
			Ref.~\cite{Yamasaki2010} and MBE-grown FeSe films of
			Ref.~\cite{Chen2010}. Data on the energy location of the peaks in PLD-grown FeSe films provided in
			Ref.~\cite{Chaluvadi2021}.} \label{tab:comparisson}
	\end{table}

\begin{figure}[ttt]
		\centering
		\includegraphics[width=0.7\linewidth]{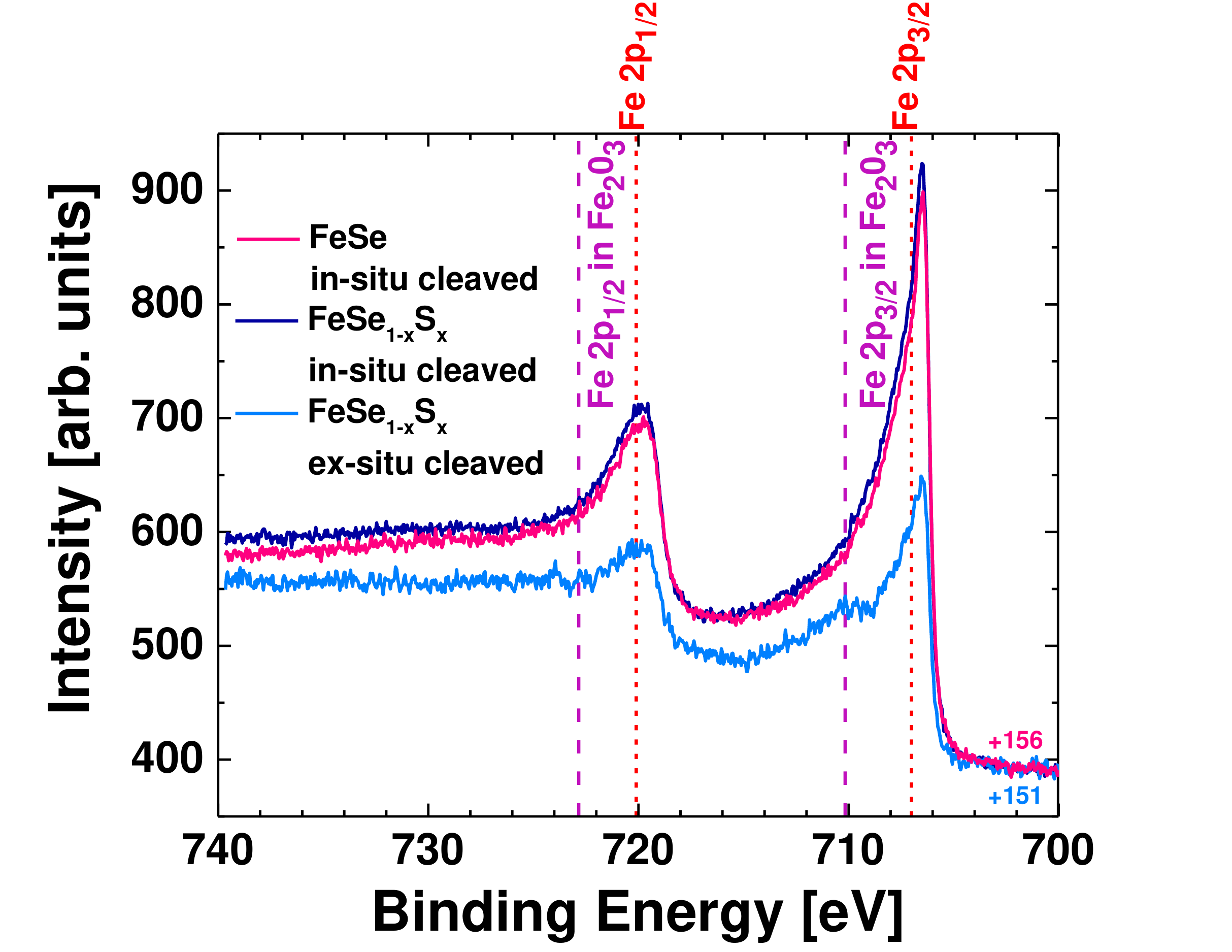}
		\caption{Detail of the Fe 2p peaks in the XPS spectra of
			FeSe$_{1-x}$S$_{x}$ (blue in-situ and turquoise ex-situ cleaved) and
			FeSe (pink in-situ cleaved) samples. For the in-situ cleaved samples
			the Fe 2p$_{3/2}$ and Fe 2p$_{1/2}$ peaks are detected at 706.5 and 719.7\,eV,
			respectively. This energy is shifted by $\sim 500$\,meV from the expected core-level energy values for the pure elements.
			In the case of the ex-situ cleaved surface, two satellite peaks located
			respectively at larger binding energies than the higher ones are also
			observed, see purple dashed lines.}
		\label{figure5}
	\end{figure}

	The differences between the spectra collected in the in-situ and
	ex-situ cleaved samples are evident in Fig.\,\ref{figure5} showing
	the energy range of the XPS spectra for the  Fe 2p core-levels. A
	clear shoulder at the left of the Fe 2p$_{3/2}$ peak and a faint one
	at the left of the Fe 2p$_{1/2}$ peak are observed in the ex-situ
	cleaved FeSe$_{1-x}$S$_{x}$ surface. These shoulders appear at
	energies corresponding to the Fe 2p core levels tabulated for the
	oxidized compound Fe$_{2}$O$_{3}$,~\cite{Moulder1992} see dashed purple lines in the
	figure. This indicates that the spectra in ex-situ cleaved samples
	show extra peaks in comparison to data in in-situ cleaved samples,
	resulting from surface oxidation of the sample. In addition, for the
	case of both, pure and S-doped FeSe samples,  notably asymmetric and
	sharp peaks are observed at 706.5 and 720\,eV, roughly corresponding
	to 500\,meV smaller energy values than the tabulated  Fe 2p$_{3/2}$
	and Fe 2p$_{1/2}$ levels of metallic Fe. This energy shift is a
	manifestation of the hybridization of the conducting Fe 3d levels
	affecting also the inner core levels, a finding previously suggested
	by DFT calculations~\cite{Subedi2008,Singh2009} and ultraviolet
	photoelectron spectroscopy data close to the Fermi level in pure FeSe.~\cite{Tsukada2011}   The  line
	shape and energy location of the Fe 2p peaks in our XPS spectra for
	in-situ cleaved surfaces are very similar to those measured in
	in-situ grown or Ar-sputtered surfaces of FeSe
	films.~\cite{Wu2007,Chen2010,Tsukada2011,Qi2011,Telesca2012,Chaluvadi2021}
	Peaks in the ex-situ cleaved sample are in contrast more rounded, as
	observed in data for ex-situ grown FeSe
	films.~\cite{Feng2004,Wu2007,Chen2010,Qi2011} In the latter case the Fe 2p$_{3/2}$ peak is generally detected around
	710\,eV, the energy expected for the Fe$_{2}$O$_{3}$
	oxide.~\cite{Feng2004,Wu2007,Chen2010} Thus, these spectral line
	shape and energy shifting of $\sim 3$\,eV for the Fe 2p levels are
	strongly affected by the sample surface preparation. The shift of
	XPS peaks due to spurious surface effects is an important issue to
	avoid in order to get information on the bulk electronic states of a
	sample since usually peak-shifts are expected due to hybridization.
	We recall that the doping of FeSe with S with $x=0.03(0.01)$ does not produce
	an extra energy shift of the Fe 2p core levels.

\begin{figure}[ttt]
		\centering
		\includegraphics[width=0.7\linewidth]{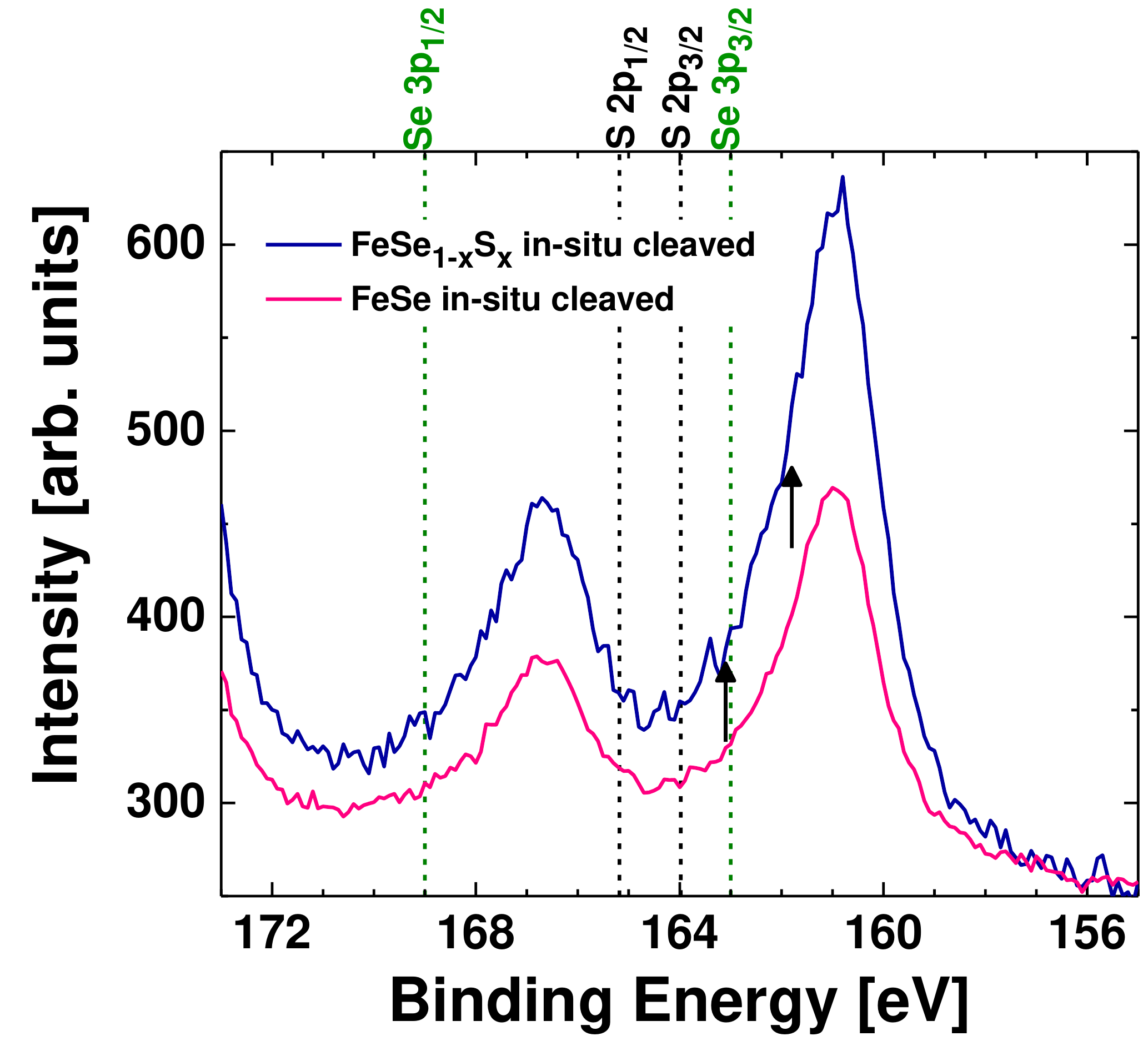}
		\caption{XPS spectra of FeSe (pink) and FeSe$_{1-x}$S$_{x}$ (blue) in-situ cleaved samples in the range of the Se 3p and S 2p levels. Dashed  lines indicate tabulated values for the Se 3p$_{1/2}$, Se 3p$_{3/2}$, S 2p$_{1/2}$ and S 2p$_{3/2}$ energy core-levels of the pure elements. Arrows indicate the energy location of the S 2p peaks in a FeS crystal according to Ref.~\cite{Liu2017}.}
		\label{figure6}
	\end{figure}

	Figure \ref{figure6} shows the XPS spectra in the energy region of
	the Se 3p levels for the in-situ cleaved pure and S-doped FeSe 
	samples. Two broad peaks centered at 160.9 and 166.6\,eV are
	observed in both samples. According to the tabulated values for the
	core levels of pure elements,~\cite{Moulder1992}  these peaks can be
	associated with the Se 3p doublet with a spin-orbit splitting $\Delta
	E \sim 6$\,eV. Moreover, for both samples the observed peaks are
	shifted 2.1\,eV for the Se 3p$_{3/2}$ peak and 2.4\,eV for the Se
	3p$_{1/2}$ peak. As in the case of the Fe 2p core levels, the energy
	location of the Se 3p peaks  is not affected by a low doping level
	of S within our experimental resolution. The black dashed lines in
	Fig.\,\ref{figure6} are located at the tabulated energy for the S 2p
	doublet core levels for pure elements. No local peaks above the
	noise level are detected at these energies nor where these peaks are
	reported for FeS single crystals,~\cite{Lai2015} see arrows in the
	figure. These peaks might be difficult to detect due to, first, their
	expected location in a flank between the  Se 3p peaks hindering the
	development of a faint peak due to the small amount of S in
	the S-doped FeSe sample. A second reason is that the photoemission cross-section of S 2p is 3 times smaller than that of Se 3p.

\begin{figure*}[hhh]
	\centering
	\includegraphics[width=0.49\linewidth]{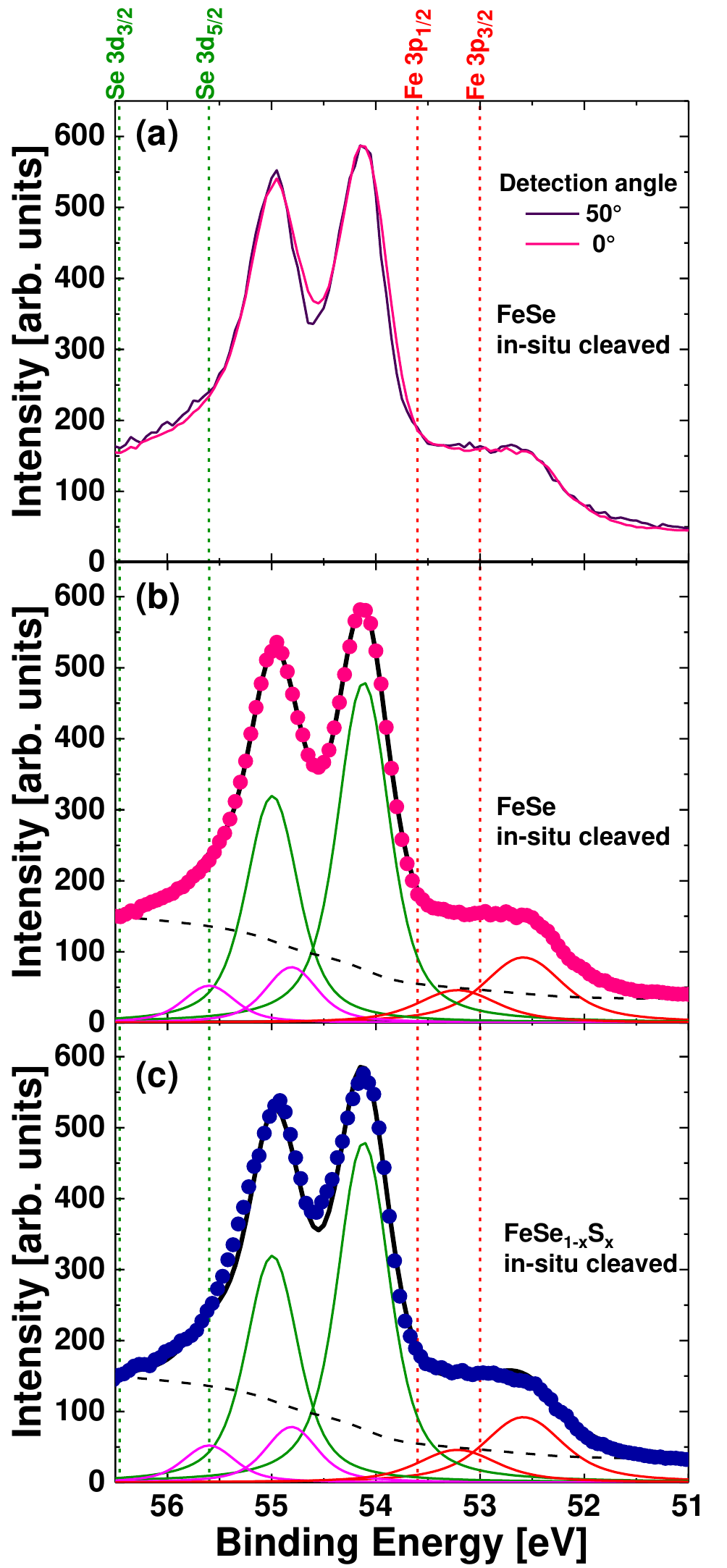}
	\caption{(a) XPS spectra in the energy region of the Se 3d and Fe3p peaks for a pure FeSe sample. Measurements performed at two different detection angles of photoemited electrons. (b)  Pink points: same data than in panel (a) for a detection angle of 0\,$^{\circ}$ (normal incidence). Full black line: Fit of the experimental data with a convolution of doublets of Voigt-like peaks after subtraction of a Shirley background (see black dashed line). A broad  peak observed around 53\,eV, close to the tabulated energy for the Fe 3p core levels, is fitted with a spin-orbit doublet (see red lines). Two pairs of peaks shown with green and magenta lines come from a principal and a minor doublet contribution associated with the Se 3d core levels. (c) Same analysis than in panel (b) for the XPS spectra measured in the S-doped crystal at normal incidence. In all panels dashed vertical lines correspond to the tabulated values for the core level energies in the pure elements. }
	\label{figure7}
\end{figure*}

	Figure \ref{figure7}\, presents a detail the XPS spectra obtained for our pure
	and S-doped FeSe samples at the energy interval of 50-60\,eV. This energy range comprises the location of the Se 3d peaks that are the highest ones, relative to the background, detected in the whole XPS spectra. For this reason, we study the spectra in this energy range to ascertain both, if there is a surface contribution, and to describe the electronic states via fits of the data. Figure \ref{figure7}\,(a) presents the XPS spectra obtained for pure
	FeSe samples for two different detection angles of photoemited
	electrons. The shape of the
	measured spectra is almost independent of the detection angle within the
	noise level. This is a very important piece of information since it implies that the effect of the peak associated with the  last-atomic-layer is negligible in our measured XPS spectra. Figures
	\ref{figure7}\,(b) and (c) show XPS spectra measured at normal
	emission  for in-situ cleaved pure and S-doped FeSe crystals,
	respectively. The spectral shape in this energy range for both types
	of samples are alike and present no noticeable quantitative
	differences: Two sharp peaks corresponding  to the Se 3d$_{5/2}$ and
	3d$_{3/2}$ core levels and a broad peak  around 53\,eV associated with
	the Fe 3p$_{3/2}$ and Fe 3p$_{1/2}$ core levels are observed as a
	shoulder in the low-enery flank of the Se 3d peaks.

	Previously published fitting of XPS data in pure FeSe samples for this energy
	range do not clearly discuss the Fe 3p peak
	contribution.~\cite{Chen2010,Yamasaki2010,Telesca2012,Chaluvadi2021}
	In view of the good quality of our XPS data, we study this
	contribution and fitted the spectra of Figs.\,\ref{figure7}\,(b)
	and (c) considering pairs of Voigt-like peaks after subtraction of a
	fitted Shirley background associated to the photoemission process of secondary electrons (see black dashed line). The corresponding
	spin-orbit splitting and statistical intensity ratios theoretically expected for both 
	peaks of the doublet were left fixed in the fits. We found that a single spin-orbit doublet
	describes the  Fe 3p levels detected at around 53\,eV, see red full
	lines. Interestingly, the Se 3d levels are properly fitted only if
	considering two spin-orbit doublets indicated with green and magenta
	full lines. These two contributions have the same expected spin-orbit splitting and statistical intensity ratio for Se 3d levels but the energy location of the peaks are left free in the fit. The main component
	corresponding to the green spin-orbit doublet is centered at exactly
	the same energies where the peaks are detected in the
	experimental spectra. This component represents 86\,\% (85\,\%) of the
	area under the curve of the Se 3d peaks fit in pure (S-doped) samples. The second contribution (magenta lines)  is shifted 0.64\,eV towards larger binding energies
	with respect to the experimentally detected peaks. The consideration
	of these minor contribution  with an area under the curve of $14$\,\% ($ 15$\,\%)  is compulsory in
	order to properly fit the data in both,  pure and S-doped FeSe
	samples.

\clearpage

	\section*{Discussion}
	
	The detection of quantitatively similar second contributions in the
	fits of the spectral shape of the Se 3d peaks in pure and S-doped
	crystals indicates that a significant amount of Se atoms has a
	different electronic environment than the rest. Since curves obtained in measurements 
	performed at different detection angles are rather coincident, 
	we rule out the presence of a surface peak in the Se 3d XPS spectrum. 
	Also, since we are studying single crystals, no interface-induced changes
	in the electronic environment of some atoms is expected as for instance detected
	in amorphous films using XPS.~\cite{Guillamon2008} Therefore, in our case the detection of the second component is necessarily associated with local variations of charge transfer that
	occur in the bulk of the samples (1\,nm thickness). These variations are very unlikely due to chemical inhomogeneities due to intergrowth of spurious phases: Not only X-ray
	diffraction but also XPS data show no detectable trace of the
	non-superconducting hexagonal phase in our high quality crystals
	that was reported in other samples.~\cite{Yamasaki2010} We can
	support this statement considering that the energy location of the
	Se 3d peaks in samples of the hexagonal non-superconducting phase
	are shifted  towards smaller binding energies, roughly 300\,meV,
	with respect to the position in the tetragonal FeSe phase. In
	contrast, the second component of the Se 3d XPS peaks  are centered
	at larger binding energies than the maximum in the experimental data that
	coincides with the center of the principal component. Second,   our
	samples present sharp Fe 2p peaks in contrast to the rounded ones
	that are detected in samples of the hexagonal phase,  in addition
	are shifted  1\,eV towards larger binding energies. Thus, the second
	component is very unlikely to have its origin in chemical
	inhomogeneities due to the inclusion of spurious phases in the sample.

	Putting the results of the fits of XPS spectra in context with the
	atomic resolution images of FeSe$_{1-x}$S$_{x}$, see
	Fig.\,\ref{fig:figure3}, we recall that  crystal disorder at
	atomic scale is more important in S-doped samples than in pure ones.
	Nevertheless, the local charge transfer accounting for the second
	contribution does not originate in the  atomic defects associated with
	S-doping since practically the same contributions to the fit are
	found in pure and S-doped samples. Thus, this type of atomic defects
	do not significantly affect the bulk electronic core levels of
	FeSe$_{1-x}$S$_{x}$ for a low doping level of $x \sim 0.03$. Our STM
	topographic images, as well as data from other
	authors,~\cite{Hanaguri2018,Huang2016,Putilov2019} do not present
	evidence of a long-wavelength modulation of the topography that
	could be eventually associated with a spatial modulation of the
	chalcogen height $z_{\rm Se}$.

	Nevertheless, both types of samples
	present Fe vacancy-induced dumbbell defects  detected by STM  that one can reasonably assume are also  present in
	all the FeSe planes probed by XPS. Every dumbbell defect entails changes
	in the electronic environment of the 4 Se atoms surrounding the Fe
	vacancy: According to DFT simulations of a FeSe monolayer they
	present a more extended electronic cloud.~\cite{Huang2016} Thus, this
	particular type of atomic defect have a noticeable impact in the
	electronic structure of FeSe$_{1-x}$S$_{x}$ and we propose that is
	at the origin of the second component detected in XPS spectra. The location
	of the doublet of the second component in larger binding energies than the
	first one is in agreement with a small fraction of Se atoms having an electronic
	environment with less charge, quite likely induced by the smaller amount of hybridization produced by the lacking Fe atom. As
	for the relation between the density of Se atoms whose electronic
	cloud is affected ($\sim 4\%$ according to STM data) and the area
	under the curve of the second component of the fit (14-15\,\%), we would
	like to stress that it is possible that the Fe vacancy of a dumbbell
	affects the orbitals of more than 4 Se surrounding atoms. Further
	DFT calculations appropriately considering the van der Waals
	interaction and the magnetic state of the Fe atom in this
	compound~\cite{Lochner2021} are important
	to quantitatively asses the possibility of the electronic cloud of
	more than 4 Se atoms and the $z_{\rm Se}$ being affected by the Fe
	vacancy producing the dumbbell defect. In addition, since the density of Se atoms entailed in dumbbell defects observed in topographies is obtained from statistics in a sample area of tens of nm$^2$, whereas XPS is probing the whole area of the sample, this difference can also come from the regions of the sample not revealed by STM having a larger concentration of dumbbell.

	\section*{Conclusions}
	
	In conclusion, we show that STM-revealed local atomic-scale defects in the crystal structure
	of the simplest Fe-based superconductor have a noticeable impact in the electronic
	properties of the material. Indeed, a minor second component, that we argue is associated with the modification of the electronic cloud of atoms surrounding these defects, is required to properly fit the
	spectral shape of XPS data. Our work paves the way for future studies trying to describe the electronic properties of Fe-based superconductors by combining the atomic-scale detection of crystal deformations and the analysis of core level states via XPS probing few atomic layers. Ultimately, the results we report here on the impact of atomic defects in the
	binding energy and spectral shape of the core levels in
	FeSe$_{1-x}$S$_{x}$ highlights the subtle interplay between the
	crystal structure and the electronic states in Fe-based
	superconductors.

	\vspace{1cm}
	
\section*{Acknowledgments}
	
	We thank J. Puig, P. Pedrazzini, E. Mart\'{i}nez and W. Sch\"{o}fferhofer for insightful discussions. Work supported by the
	Argentinean ANPCyT through grants PICT
	2017-2182 and 2018-1533, and by the Universidad Nacional de Cuyo
	research grants 06/C566 and 06/C575. Y. F. thanks funding from the
	Alexander von Humboldt Foundation through the Georg Forster Research
	Award.

\end{document}